# Implications and consequences of ferromagnetism universally exhibited by inorganic nanoparticles

A. Sundaresan, C. N. R. Rao

Chemistry and Physics of Materials Unit, New Chemistry Unit, and International Centre for Materials Science, Jawaharlal Nehru Centre for Advanced Scientific Research, Jakkur P.O., Bangalore 560 064 India

Occurrence of surface ferromagnetism in inorganic nanoprticles as a universal property not only explains many of the unusual magnetic features of oxidic thin films, but also suggests its possible use in creating new materials, as exemplified by multiferroic $BaTiO_3$ nanoparticles. While the use of Mn-doped ZnO and such materials in spintronics appears doubtful, it is possible to have materials exhibiting coexistence of (bulk) superconductivity with (surface) ferromagnetism.

# 1. Background

Thin films of nonmagnetic oxides such as $HfO_2$, $TiO_2$, $In_2O_3$ were surprisingly found to exhibit ferromagnetism at room temperature a few years ago [1-3]. The origin of magnetism in these films was not entirely clear till recently. However, recent studies on inorganic nanoparticles have shown room-temperature ferromagnetism to be a universal characteristic of these nanomaterials, typical materials exhibiting ferromagnetism being ZnO, $Al_2O_3$, MgO, GaN, and CdS, all of which are diamagnetic in bulk form [4-7]. In Fig. 1, we show typical magnetization versus field curves in the case of ZnO and GaN nanoparticles and of the corresponding bulk samples at 300 K to illustrate the unique characteristic of nanoparticles. It is obvious from this figure that the nanoparticles exhibit magnetic hysteresis at room-temperature and above, although the bulk samples show diamagnetic behaviour expected of band insulators. The Curie temperatures of these nanoparticles turn out to be well above 400 K which is the high temperature limit of most magnetometers. Table I lists the values of saturation magnetization, coercive field and remanence at 300 K for a number of inorganic nanoparticles studied recently [4-7]. The saturation magnetization is generally low, five orders less than that of iron. However, the observed ferromagnetism is intrinsic to the nanoparticles and is not due to any magnetic impurity phase or any other artefact. The values of saturation magnetization, $M_S$, in Table I are comparable to those reported for a number of wide-band gap oxide semiconductors substituted with a few percent of transition metal ions [8-11]. Thus, Mn doped ZnO and GaN are reported to show weak magnetic hysteresis at 300 K with similar $M_S$ values [8, 12]. The universality associated with the ferromagnetism of inorganic nanoparticles has certain important implications and consequences. We discuss these aspects in this article. Before doing so, we examine the origin of ferromagnetism in nanoparticles.

## 2. Origin of ferromagnetism in thin films and nanoparticles of inorganic materials

It has been known for many years that neutral cation vacancies in simple binary oxides can give rise to holes on neighbouring oxygen ions which can give rise to a triplet ground state [13]. In fact, electron paramagnetic resonance (EPR) studies on γ-ray irradiated CaO single crystals have shown the presence of two holes trapped by a single neutral cation vacancy [14]. Based on *ab initio* band structure calculations, it has been shown that a Ca vacancy concentration of 3% can lead to a ferromagnetic ground state [15]. First-principles total energy calculations have shown that it requires a minimum of 4.9% of neutral Ca vacancies with ferromagnetic interaction extending to four neighbours [16]. These calculations also suggest a nonequilibrium growth process to be responsible for such a high vacancy concentration. Experimentally, thin films of nonmagnetic $HfO_2$ are known to exhibit ferromagnetism at room temperature [1]. This was explained on the basis of density functional theoretical calculations as due to Hf vacancies giving rise to a ferromagnetic ground state [17]. Room-temperature ferromagnetism in thin films of nonmagnetic oxides such as $TiO_2$, $In_2O_3$ and ZnO is generally explained invoking a nonequilibrium growth process which leads to high defect concentrations [2, 18, 19]. Growth conditions greatly influence the defect concentration in thin films and hence the magnetism. It is for this reason, that the results on $HfO_2$ were not readily reproducible [20].

Unlike thin films, nanoparticles can have a large number of defects at the surface of the particles, the concentration of defects being controlled by the size of the particles. In fact, it has been shown that smaller nanoparticles exhibit room-temperature ferromagnetism whereas the bigger ones show diamagnetism [7]. This demonstrates that ferromagnetism in the inorganic nanoparticles could arise from surface defects. The nature of defects (cation or anion vacancies) could vary from one material to another. For example, the origin of magnetism in MgO is associated with cation defects whereas it is due to oxygen vacancies in $BaTiO_3$ [21, 22]. In the case of GaN, Ga vacancies are responsible for the magnetism [23].

It is possible to relate surface defects to ferromagnetism in as-prepared nanoparticles of ZnO. ZnO exhibits a photoluminescence band in the visible region centered around 560 nm due to defects besides the one at 380 nm intrinsic to the material [24]. The intensity of the defect band decreases on annealing ZnO nanoparticles in oxygen. It is noteworthy that the decrease in the defect PL band intensity parallels the change in the magnetic properties of ZnO nanoparticles.

## 3. Magnetism in dilute magnetic semiconductors

Dilute magnetic semiconductors (DMS) are wide band gap semiconducting materials substituted with few percent of transition metal ions having local magnetic moments. While explaining ferromagnetism in Mn substituted GaAs [25], Dietl *et al* [26] proposed that p-type GaN and ZnO can be ferromagnetic with Tc greater than 300 K depending upon the concentration of holes and magnetic ions substituted. Following this prediction, room-temperature ferromagnetism was reported in Co doped $TiO_2$ thin films [27]. The trouble with Co doping is that Co ions can readily get reduced to Co metallic particles which cannot be detected by standard experimental techniques when present in small concentration. Room-temperature ferromagnetism has been reported in thin films and bulk samples of Mn substituted ZnO by few workers [8, 28], but several studies on well-characterized bulk samples of ZnO substituted with Mn did not show collective magnetism [29]. In particular, samples of Mn doped ZnO prepared at low temperatures have failed to exhibit ferromagnetism [30, 31]. The values of saturation magnetization reported for thin films and nanoparticles of Mn substituted ZnO are around $1- 8 \times 10^{-3}$ emu/g. This value of $M_S$ is comparable to the values given in Table I for a variety of undoped nanoparticles. Furthermore, the Curie temperature is higher than 400 K. These results suggest that the origin of magnetism in the transition metal-doped oxide semiconductors is likely to be due to point defects which can occur in large concentrations in thin films and at the surface of the nanoparticles due to low defect formation energies [32]. These arguments are likely to be valid in the case of Mn-doped GaN as well, although $Mn^{2+}$ in place of Ga can introduce some carriers. The above discussion clearly indicates how the magnetism observed in dilute magnetic semiconductors such as Mn doped ZnO would not be suitable for use in spintronics [11]. Magnetism of inorganic nanoparticles may have, however, other applications as typified by multiferroic $BaTiO_3$ as discussed below.

## 4. Nanocrystalline $BaTiO_3$, a new kind of multiferroic

Multiferroic materials having two or more ferroic properties received much attention in the recent years because of a possibility of new device applications envisaged from a strong coupling between the order parameters [33, 34]. Such multiferroic materials are rare since the ferroelectricity in oxide materials arises from the hybridization between the empty d-orbitals of transition metals and the oxygen $O_{2p}$ orbital. Occupation of d-orbitals doest not favour ferroelectricity but is essential for magnetism. These two properties are therefore antagonistic [35]. but recent research has demonstrated that there can be novel alternative routes to multiferroic materials [36]. Perovskite oxides of the type $BiMO_3$ (M=Fe & Mn) have been shown to exhibit multiferroic properties due to the stereochemical activity of lone pair of electrons associated with $Bi^{3+}$ ions [37]. Rare earth manganites $RMnO_3$ with R=rear earths having relatively smaller ionic size show ferroelectricity due to the distortion of $MnO_5$ polyhedra and an antiferromagnetic ordering below 100 K [38]. Spiral magnetic ordering also gives rise to ferroelectricity as in $TbMnO_3$ [39] and $CoCr_2O_4$ [40]. Charge-ordered systems such as $LuFe_2O_4$ [41, 42] and $R_{1-x}Ca_xMnO_3$ systems have also been shown to exhibit ferroelectricity due to combined site-centered and bond-centered ordering [43, 44]. It has been possible to exploit surface magnetism in nanoparticles of $BaTiO_3$ along with the ferroelectricity of the bulk material to render it multferroic. Shown in Fig. 2 are the magnetization and polarization hysteresis curves of $BaTiO_3$ nanoparticles [22]. Interestingly, there is coupling between the surface ferromagnetism and core ferroelectricity as evidenced from the magntocapacitance effect.

## 5. Ferromagnetic superconductors

Superconductivity and magnetism were generally considered to be incompatible until the discovery of coexistence of ferromagnetism and superconductivity in $UGe_2$ [45] and $ZrZn_2$ [46]. This has created considerable interest in exploring the coexistence of magnetism and superconductivity. Since ferromagnetism in the nanoparticles of the otherwise nonmagnetic materials arises from surface defects with the core of the particles remaining diamagnetic, we thought that the surface ferromagnetism could coexist with bulk superconductivity. We, therefore, prepared an intrinsic core shell structure with a superconducting core and a ferromagnetic shell [5]. In Fig. 3, we show the room-temperature magnetic hysteresis curves for nanoparticles of the high-$T_c$ superconductor $YBa_2Cu_3O_{7-y}$ (YBCO) at 300 and 91 K. It is to be mentioned that the Curie temperature of YBCO nanoparticles is around 800 K [7]. The magnetization curve for bulk YBCO at 300 K shows weak paramagnetic behaviour typical of Pauli paramagnetism. The shape of the magnetization curve at 90 K suggests possible coexistence of superconductivity at the core and surface ferromagnetism.

## 6. Outlook

The discussion in the previous sections should suffice to indicate the important significance of the observation of ferromagnetism in nanoparticles of inorganic materials which are normally nonmagnetic. While the occurrence of ferromagnetism in nanoparticles of ZnO and GaN casts serious doubts regarding the use of Mn doping in these materials for spintronics, it seems entirely possible that the surface magnetism of nanoparticles can be suitably exploited for certain novel applications.

**Figure Captions**

Fig. 1. (a) ZnO and (b) GaN nanoparticles showing ferromagnetic hysteresis at room-temperature. The bulk ZnO and GaN exhibit diamagnetic behaviour.

Fig. 2. (a) Magnetic and (b) ferroelectric hysteresis curves exhibited by nanocrystalline BaTiO$_3$ at room temperature. Bulk BTO is diamagnetic and ferroelectric.

Fig. 3. Magnetization curves of superconducting YBCO nanoparticles showing surface ferromagnetism.[5]

Table 1 Magnetic properties of inorganic nanoparticles

| Materials | Average diameter of Nanoparticles | Saturation magnetization (300 K) (emu/g) at H = 5 kOe | Remanence (emu/g) | Coercive field (Oe) |
|---|---|---|---|---|
| CeO$_2$ | 15 nm | $1.87 \times 10^{-3}$ | $4.0 \times 10^{-5}$ | 22 |
| Al$_2$O$_3$ | 4 nm | $4.79 \times 10^{-3}$ | $4.6 \times 10^{-4}$ | 76 |
| ZnO | 30 nm | $4.54 \times 10^{-4}$  H= 2 kOe | $2.3 \times 10^{-4}$ | 240 |
| In$_2$O$_3$ | 12 nm | $3.7 \times 10^{-4}$  H= 2 kOe | $8.9 \times 10^{-5}$ | 76 |
| NbN | nm | $1.50 \times 10^{-3}$ | $3.5 \times 10^{-4}$ | 138 |
| CdS | nm | $3.77 \times 10^{-3}$ | $3.8 \times 10^{-4}$ | 75 |
| GaN | nm | $1.25 \times 10^{-3}$ | $2.0 \times 10^{-4}$ | 136 |

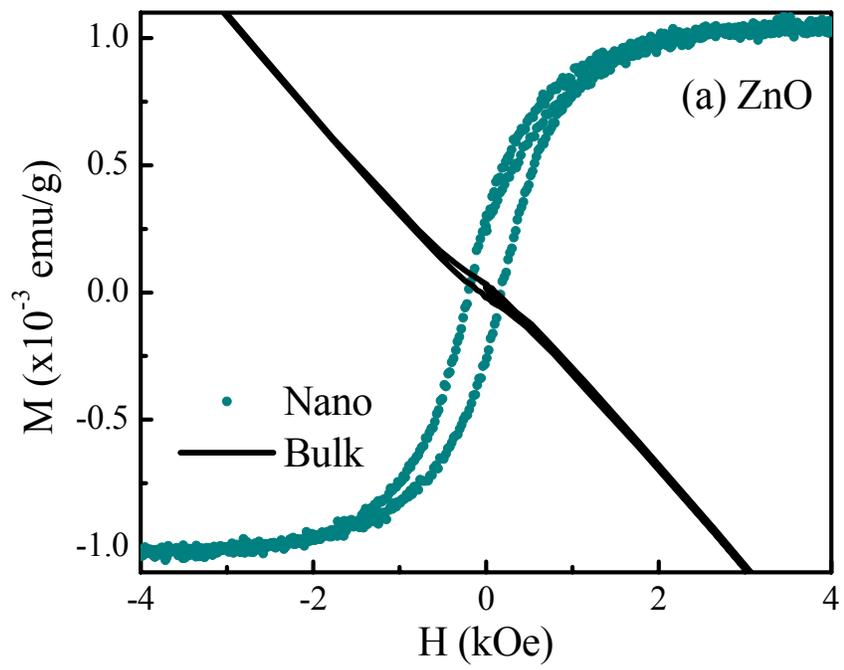

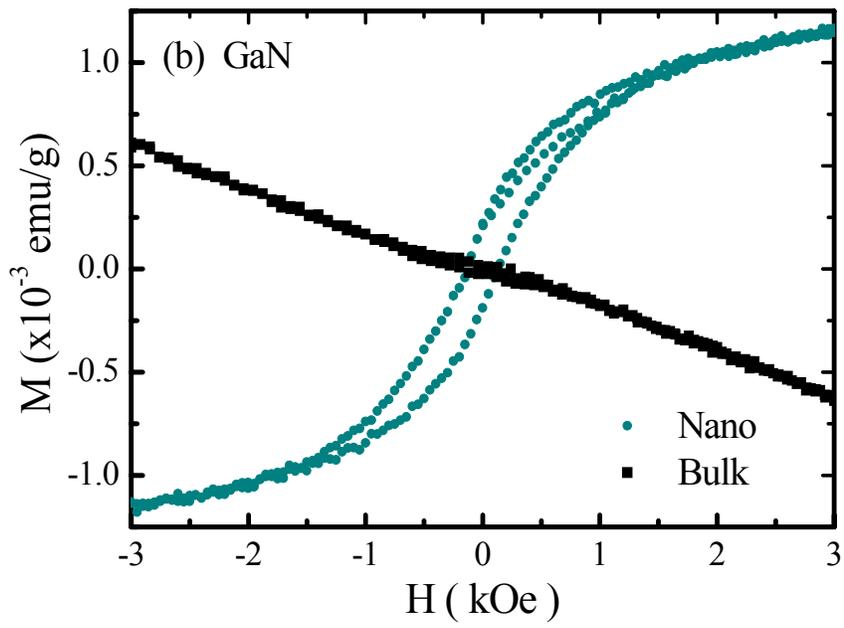

Fig. 1

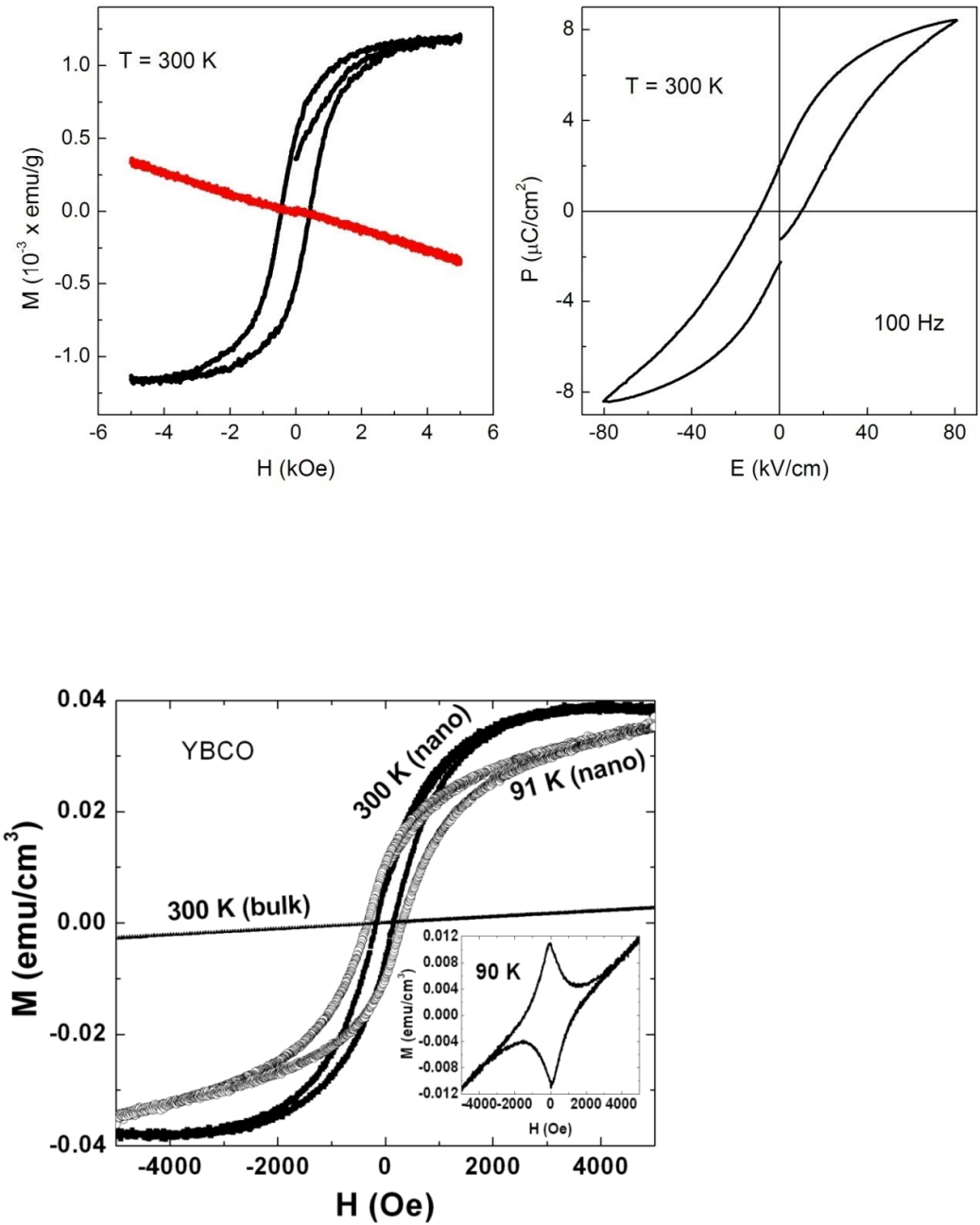

Fig. 3